\documentclass[aps,prd,preprintnumbers,superscriptaddress,nofootinbib,floatfix,twocolumn,notitlepage,amsmath,amssymb]{revtex4-2}

\usepackage{graphicx}
\usepackage{dcolumn}
\usepackage{xspace}
\usepackage[utf8]{inputenc}
\usepackage{float}
\usepackage{subfigure}
\usepackage{url}
\usepackage{color}
\usepackage[dvipsnames,table]{xcolor}
\usepackage{comment}
\usepackage{adjustbox}
\usepackage{booktabs} 
\usepackage{colortbl} 

\def\beq#1\eeq{\begin{align}#1\end{align}}

\newcommand{\eq}[1]{Eq.~(\ref{#1})}

\newcommand{\Bbar}{\,\overline{\!B}}

\RequirePackage{xspace}
\def\Bbar    {\kern 0.18em\overline{\kern -0.18em B}{}\xspace}

\def\Jpsi   {J\mbox{\footnotesize$\!/\psi$}}

\definecolor{BlueViolet}{rgb}{0.2, 0.00, 0.7}
\definecolor{Blue}{rgb}{0.15, 0.00, 0.9}
\definecolor{halayaube}{rgb}{0.4, 0.22, 0.33}
\definecolor{sanddune}{rgb}{0.59, 0.44, 0.09}

\usepackage[colorlinks=true,linkcolor=Blue,citecolor=Blue,urlcolor=BlueViolet]{hyperref}
\usepackage[hyphenbreaks]{breakurl} 

\begin{document} 

\preprint{PSI-PR-22-34}
\preprint{ZU-TH 56/22}
\preprint{TTP22-069}
\preprint{P3H-22-113}
\preprint{KEK-TH-2474}
\title{\boldmath 
Impact of \texorpdfstring{$\Lambda_b\to \Lambda_c\tau\nu$}{Lambdab to Lambdac tau nu} measurement on New Physics in \texorpdfstring{$b\to c \, l \nu$}{b to c  l nu} transitions
}

\author{Marco Fedele}
\email[]{marco.fedele@kit.edu}
\affiliation{Institut f\"ur Theoretische Teilchenphysik (TTP), Karlsruhe Institute of Technology, D-76131 Karlsruhe, Germany}

\author{Monika Blanke} 
\email[]{monika.blanke@kit.edu}
\affiliation{Institut f\"ur Theoretische Teilchenphysik (TTP), Karlsruhe Institute of Technology, D-76131 Karlsruhe, Germany}
\affiliation{Institut f\"ur Astroteilchenphysik (IAP),
Karlsruhe Institute of Technology, D-76344 Eggenstein-Leopoldshafen, Germany}

\author{Andreas Crivellin} 
\email[]{andreas.crivellin@cern.ch}
\affiliation{Paul Scherrer Institut, CH–5232 Villigen PSI, Switzerland}
\affiliation{Physik-Institut, Universit\"at Z\"urich, Winterthurerstrasse 190, 8057 Z\"urich, Switzerland}

\author{Syuhei Iguro}
\email[]{igurosyuhei@gmail.com}
\affiliation{Institut f\"ur Theoretische Teilchenphysik (TTP), Karlsruhe Institute of Technology, D-76131 Karlsruhe, Germany}
\affiliation{Institut f\"ur Astroteilchenphysik (IAP),
Karlsruhe Institute of Technology, D-76344 Eggenstein-Leopoldshafen, Germany}

\author{Teppei Kitahara} 
\email[]{teppeik@kmi.nagoya-u.ac.jp}
\affiliation{Institute for Advanced Research \& Kobayashi-Maskawa Institute for the Origin of Particles and the Universe, 
Nagoya University,  Nagoya 464--8602, Japan}
\affiliation{KEK Theory Center, IPNS, KEK, Tsukuba 305-0801, Japan}
\affiliation{CAS Key Laboratory of Theoretical Physics, Institute of Theoretical Physics, Chinese Academy of Sciences, Beijing 100190, China}

\author{Ulrich Nierste}
\email[]{ulrich.nierste@kit.edu}
\affiliation{Institut f\"ur Theoretische Teilchenphysik (TTP), Karlsruhe Institute of Technology, D-76131 Karlsruhe, Germany}

\author{Ryoutaro Watanabe}
\email[]{wryou1985@gmail.com}
\affiliation{INFN, Sezione di Pisa, Largo B. Pontecorvo 3, 56127 Pisa, Italy}


\begin{abstract}
\noindent
Measurements of the branching ratios of $B \to D^{(*)}\tau\bar\nu/B \to D^{(*)}\ell\bar\nu$ and $B_c\to \Jpsi\, \tau\bar\nu/B_c\to \Jpsi\, \ell\bar\nu$ by the BaBar, Belle and LHCb collaborations consistently point towards an abundance of taus compared to channels with light leptons. However, the ratio $\Lambda_b \to\Lambda_c \tau\bar\nu/\Lambda_b \to\Lambda_c \ell\bar\nu$ shows a relative deficit in taus. In this paper, we critically address whether data still points towards a coherent pattern of deviations, in particular in light of the sum rule relating these decays in a model-independent way. We find that no common new physics explanation of all ratios is possible (within $2\sigma$ or $1.5\sigma$, depending on the ${\cal R}(\Lambda_c)$ normalization to light lepton channels). While this inconsistency could be a statistical fluctuation, further measurements are required in order to converge to a coherent pattern of experimental results.
\end{abstract}

\maketitle


\section{Introduction}

The Standard Model (SM) has a solid experimental foundation since its formulation half a century ago~\cite{Glashow:1961tr,Weinberg:1967tq,Salam:1968rm}. However, several incontestable observations, like the presence of dark matter or neutrino oscillation (see, e.g., Refs.~\cite{Arbey:2021gdg,Gonzalez-Garcia:2022pbf} for recent reviews), prove the existence of New Physics (NP). A useful approach for searching for NP is to look at the violation of (approximate) symmetries of the SM, like, e.g., lepton flavor universality (LFU) which is only broken in the SM Lagrangian by the small Yukawa couplings.

In fact, several hints for the violation of LFU have emerged over the last years (see e.g.~Refs.~\cite{Bernlochner:2021vlv,Fischer:2021sqw,London:2021lfn,Crivellin:2021sff,Crivellin:2022qcj} for a recent review). In particular, ratios of the semi-leptonic $b$ hadrons decays
\begin{align}
\begin{aligned} {\cal R}(D^{(*)})\,&\equiv\,
{\rm BR}(B\to D^{(*)} \tau \bar\nu)/{\rm BR}(B\to D^{(*)} \ell \bar\nu)\,,\\
{\cal R}(\Jpsi)\,&\equiv\,
{\rm BR}(B_c\to \Jpsi\, \tau\bar\nu)/{\rm BR}(B_c\to \Jpsi\,\ell \bar\nu)\,,\\
{\cal R}(\Lambda_c)\,&\equiv\,
{\rm BR}(\Lambda_b \to\Lambda_c \tau\bar\nu)/{\rm BR}(\Lambda_b \to\Lambda_c \ell\bar\nu)\,,
\end{aligned}
\end{align}
where CKM and hadronic uncertainties drop out and are reduced, respectively, show deviations from the SM predictions.\footnote{Here and in the following, $\ell=e, \mu$, while $l=e,\mu,\tau$.}

Both ${\cal R}(D)$ and ${\cal R}(D^*)$ have been measured by 
BaBar~\cite{BaBar:2012obs,BaBar:2013mob} and Belle~\cite{Belle:2015qfa,Belle:2016ure,Hirose:2016wfn,Hirose:2017dxl, Belle:2019rba} and a first combined measurement of these ratios has just recently been announced by LHCb~\cite{LHCbSem}, which previously had measured only the latter one~\cite{Aaij:2015yra,Aaij:2017uff,Aaij:2017deq}. A global average for these quantities has been provided by the HFLAV collaboration~\cite{HFLAV:2022pwe},
\begin{eqnarray}
\begin{aligned}
{\cal R}(D)\,=\,0.358\pm0.025\pm0.012  \,, \label{rdhflav}\\
{\cal R}(D^*)\,=\,0.285\pm0.010\pm0.008\,,
\label{Eq:HFLAV}
\end{aligned}
\end{eqnarray}
where the first uncertainty is statistical and the second is systematic. When comparing this result with the recent SM predictions~\cite{HFLAV:2022pwe,Bigi:2016mdz,Bernlochner:2017jka,Jaiswal:2017rve,Gambino:2019sif,Bordone:2019vic,Martinelli:2021onb},
\begin{eqnarray}
\begin{aligned}
{\cal R}_{\rm SM}(D)\,=\,0.298\pm0.004 \,, \\
{\cal R}_{\rm SM}(D^*) \,=\,0.254\pm0.005 \,,
\label{eq:HFLAVSM}
\end{aligned} 
\end{eqnarray}
one observes a tension at the level of $3.2\,\sigma$. As the determination of $|V_{cb}|$ from the modes with light leptons is consistent with global CKM fits~\cite{UTfit:2022hsi,CKMfitter}, it is regularly assumed that the deviation implies an over-abundance of taus. 

An analogous behavior has been observed for ${\cal R}(\Jpsi)$~\cite{LHCb:2017vlu}
\beq
{\cal R}(\Jpsi)=0.71 \pm 0.17 \pm 0.18\,.
\eeq
To compare this result with SM predictions, we can rely on the latest estimates~\cite{Cohen:2018dgz,Leljak:2019eyw,Harrison:2020nrv,Harrison:2020gvo}, 
\beq
{\cal R}_{\rm SM}(\Jpsi) = 0.258 \pm 0.004\,,
\eeq
that are compatible with data at the $1.8\,\sigma$ level. However, we are still missing a determination of the tensor form factors from lattice, and the lack of a precise knowledge for these form factors from other sources precludes an accurate NP analysis~\cite{Watanabe:2017mip,Tran:2018kuv,Murphy:2018sqg,Leljak:2019eyw,Colangelo:2022lpy}. For this reason, we do not include this observable in our NP  analysis.

Finally, LHCb~\cite{LHCb:2022piu} finds
\beq
{\cal R}(\Lambda_c) = 0.242 \pm 0.026 \pm 0.040 \pm 0.059\,,
\eeq
where the first uncertainty is statistical, the second is systematic and the third is due to external branching fraction measurements. A recent reanalysis of this result has been performed in Ref.~\cite{Bernlochner:2022hyz} where, in order to reduce systematic errors, the tau decay channel measured by the LHCb collaboration is normalized to the SM prediction for $\Gamma(\Lambda_b\to\Lambda_c\mu\bar{\nu})$, rather than employing its experimental average. Such a procedure improves the accuracy and slightly amplifies the central value, yielding
\beq
\mathcal{R}(\Lambda_c) = |0.04/V_{cb}|^2 (0.285 \pm 0.073)\,.
\eeq
In comparison the SM prediction, where the absence of a subleading Isgur-Wise function at $\mathcal{O}(\bar \Lambda/m_{c,b})$ in the $\Lambda_b \to \Lambda_c$ transition suppresses the theoretical uncertainty~\cite{Neubert:1993mb}, is equal to~\cite{Gutsche:2015mxa,Shivashankara:2015cta,Detmold:2015aaa,Li:2016pdv,Datta:2017aue,Bernlochner:2018kxh,Bernlochner:2018bfn}
\beq\label{eq:RLc_SM}
{\cal R}_{\rm SM}(\Lambda_c) = 0.324 \pm 0.004\,.
\eeq
Although this value does not point towards a strong tension with the SM, it actually hints this time to an under-abundance of taus.

This opposite behavior compared to the other ratios is unexpected as all processes are described by the same effective Hamiltonian for $b \to c l\nu$ transitions. Many model-independent NP  analyses have been performed to explain either the deviation observed in ${\cal R}(D)$ and ${\cal R}(D^*)$~\cite{
Fajfer:2012jt,Datta:2012qk,
Tanaka:2012nw,
Freytsis:2015qca,
Bardhan:2016uhr,Bhattacharya:2016zcw,Celis:2016azn,
Alok:2017qsi,Azatov:2018knx,Bhattacharya:2018kig,
Huang:2018nnq,Angelescu:2018tyl,
Iguro:2018fni,
Murgui:2019czp,Shi:2019gxi,Becirevic:2019tpx,
Iguro:2022yzr}, and/or ${\cal R}(\Lambda_c)$ alone~\cite{Bernlochner:2018bfn,DiSalvo:2018ngq,Ray:2018hrx,Penalva:2019rgt,Ferrillo:2019owd,Mu:2019bin,Becirevic:2022bev}, with NP effects connected to tau leptons. However, a joint description of the three LFU ratios is mandatory, because the three decay modes are correlated in a model-independent way: ${\cal R}(D)$, ${\cal R}(D^*)$ and $\mathcal{R}(\Lambda_c)$ fulfill a sum rule which is rooted in their properties in the heavy quark limit~\cite{Blanke:2018yud,Blanke:2019qrx}.

The intent of this paper is therefore to critically scrutinize the compatibility of data. We try to understand, by means of an EFT approach, whether it is possible to introduce further NP effects in order to address experimental measurements, or if on the other hand we are facing a situation where current results are incompatible among themselves. While most previous analyses were restricted to NP contributions in tau final states, we also consider the possibility to introduce NP coupled to light leptons, thereby modifying the sum rule in order to potentially accommodate data.

This paper is organized a follows: in Sec.~\ref{sec:EFT} we introduce the EFT formalism employed to perform the NP analyses and in Sec.~\ref{sec:sumrule} we update the sum rule, which is modified once taking the latest results into account. In Sec.~\ref{sec:NP} we review all the possible, simple UV completions that can produce the effects described by the EFT at the low-scale, and in Sec.~\ref{sec:res} we report the results of our fits. We draw our conclusions in Sec.~\ref{sec:concl}.

\vspace{1em}
\section{EFT formalism}\label{sec:EFT}

We use the effective Hamiltonian
\begin{equation}
\renewcommand{\arraystretch}{1.8}
\begin{array}{r}
 {\cal H}_{\rm eff}=  2\sqrt{2} G_{F} V^{}_{cb} \big[(1+C_{V_L}^l) O_{V_L}^l + C_{S_R}^l O_{S_R}^l
 \\   +C_{S_L}^l O_{S_L}^l+C_{T}^l O_{T}^l\big] \,,\quad\quad\quad
\end{array}
\label{Heff}
\end{equation}
with the dimension-six operators
\begin{equation}
\renewcommand{\arraystretch}{1.8}
\begin{array}{l}
   O_{V_L}^l  = \left(\bar c\gamma ^{\mu } P_L b\right)  \left(\bar l \gamma_{\mu } P_L \nu_{l}\right)\,, \\ 
   O_{S_R}^l  = \left( \bar c P_R b \right) \left( \bar l P_L \nu_{l}\right)\,, \\
   O_{S_L}^l  = \left( \bar c P_L b \right) \left( \bar l P_L \nu_{l}\right)\,,   \\
   O_{T}^l  = \left( \bar c \sigma^{\mu\nu}P_L  b \right) \left( \bar l \sigma_{\mu\nu} P_L \nu_{l}\right)\,,   \\
\end{array}
\label{Oeff}
\end{equation}
where $\sigma_{\mu\nu}=\frac i2 [\gamma_\mu,\gamma_\nu]$. Note that in our convention for the effective Hamiltonian the Wilson coefficients (WCs) $C^l_i$ describe a genuine NP effect, and vanish in the SM. Moreover, NP effects due to light right-handed neutrinos, like the ones considered e.g., in Refs.~\cite{Iguro:2018qzf,Asadi:2018wea,Greljo:2018ogz,Robinson:2018gza,Azatov:2018kzb,Babu:2018vrl}, or right-handed quark vector currents, which are LFU at the dimension-six level~\cite{Buchmuller:1985jz,Grzadkowski:2010es,Alonso:2014csa,Aebischer:2015fzz}, are not included in our analysis.

Finally, it is important to remember that the operators and WCs in \eq{Heff} are scale-dependent. We perform our analysis for a heavy NP scale, which we take to be 2$\,$TeV for concreteness. To connect these coefficients to the decay scale $\mu=\mu_b = 4.2\,$GeV,  we use the renormalization-group evolution (RGE) for the dimension-six operators at next-to-leading order in QCD and leading order in EW interactions,  including the top-quark threshold corrections~\cite{Gonzalez-Alonso:2017iyc} and taking the QCD one-loop matching corrections (for general scalar and vector leptoquarks whose couplings are invariant under the SM gauge group) into account at the NP scale~\cite{Aebischer:2018acj}, 
\begin{align}
C_{V_L}^l (\mu_b)&= 1.12 \, C_{V_L}^l (2\,\mbox{TeV})\,,\nonumber \label{wcrun}\\[1mm]
C_{S_R}^l(\mu_b) &=2.00\, C_{S_R}^l (2\,\mbox{TeV}) \,, \\[1mm]
  \left( \begin{array}{c}
      C_{S_L}^l(\mu_b) \\ C_T^l(\mu_b)           
   \end{array}\right) &=  \left( \begin{array}{rr}
          1.91 & -0.38 \\
           0.    & 0.89 
   \end{array}\right)
  \left( \begin{array}{c}
         C_{S_L}^l(2\,\mbox{TeV}) \\ C_T^l(2\,\mbox{TeV})    
   \end{array}\right) . \nonumber
\end{align}

\section{Updated sum rule}\label{sec:sumrule}

$\mathcal{R}(D^{(\ast)})$ and $\mathcal{R}(\Lambda_c)$ have strong theoretical correlations as they depend on the same transition at the quark level. Given the update of the form factors relevant for $B\to D^{(*)}$ transitions~\cite{Bordone:2019vic}, together with the newly measured value for $\mathcal{R}(\Lambda_c)$~\cite{LHCb:2022piu} and the updated measurements for $\mathcal{R}(D^{(\ast)})$~\cite{LHCbSem}, we update here the sum rule connecting the three LFU ratios~\cite{Blanke:2018yud,Blanke:2019qrx}. Indeed, the update of the form factors has an impact on the coefficients feeding into the sum rule, while the additional experimental information for $\mathcal{R}(D^{(\ast)})$ gives a more precise prediction for $\mathcal{R}(\Lambda_c)$, which can now be directly compared with its experimental measurement. For this we start from a semi-numerical formula for $\mathcal{R}(\Lambda_c)$, assuming NP contributions to the tau channel only and using the $\Lambda_b \to \Lambda_c$ lattice QCD results  of Refs.~\cite{Detmold:2015aaa, Datta:2017aue,Murgui:2019czp}:
\begin{widetext}
\beq
\label{eq:RLambda}
\frac{\mathcal{R}(\Lambda_c)}{\mathcal{R}_{\rm SM}(\Lambda_c)}= & \left|1+C_{V_L}^\tau\right|^2
+ 0.50 \,\textrm{Re}\left[ \left(1 +C_{V_L}^\tau\right) C_{S_R}^{\tau\ast}
\right]
 + 0.33 \,\textrm{Re}\left[ \left(1 +C_{V_L}^\tau\right) C_{S_L}^{\tau\ast} 
\right]
+ 0.52 \, \textrm{Re}\left( C_{S_L}^\tau C_{S_R}^{\tau\ast}  \right)
\nonumber \\
&
+ 0.32 \, \left(|C_{S_L}^\tau|^2 + |C_{S_R}^\tau|^2\right) 
 -3.11 \,\textrm{Re}\left[ \left(1 +C_{V_L}^\tau\right)  C_T^{\tau\ast}\right]
+ 10.4 \, |C_T^\tau|^2\,,
\eeq
where the Wilson coefficients have to be considered at the scale $\mu = \mu_b$ \cite{Datta:2017aue}, and we used $m_c (\mu_b) =  0.92\,$GeV in the computation of the form factors of the scalar and pseudoscalar currents.
\end{widetext}

Combining this with the updated general NP formulae for $\mathcal{R}(D^{(\ast)})$~\cite{Iguro:2022yzr} and Eq.~\eqref{eq:RLambda}, we find\footnote{We obtain this sum rule by imposing a condition such that a $C_{V_L}^\tau C_{S_L}^{\tau\ast}$ interference term is absent in $\delta_{\Lambda_c}$, while Ref.~\cite{Blanke:2018yud} has imposed such a condition for $C_{V_L}^\tau C_{S_R}^{\tau\ast}$. We find that although both procedures are numerically equivalent, our procedure is slightly more accurate whenever the $\delta_{\Lambda_c}$ term is ignored.}
\begin{align} \label{eq:Sumrule}
\frac{\mathcal{R}(\Lambda_c)}{\mathcal{R}_{\rm SM}(\Lambda_c)}=
0.280\, \frac{\mathcal{R}(D)}{\mathcal{R}_{\rm SM}(D)} + 0.720\,\frac{\mathcal{R}(D^\ast)}{\mathcal{R}_{\rm SM}(D^\ast)} + \delta_{\Lambda_c}\,,
\end{align}
with
\begin{align} \label{eq:deltaLambdac}
\delta_{\Lambda_c} =& \ 
\textrm{Re}\left[\left(1 +C_{V_L}^\tau \right) 
\left( 
  0.314\, C_T^{\tau\ast} - 0.003\,   C_{S_R}^{\tau\ast}
\right)\right] \nonumber\\
& +0.014\,\left( |C_{S_L}^\tau|^2 + |C_{S_R}^\tau|^2\right) \nonumber\\
&+ 0.004 \,\textrm{Re}\left( C_{S_L}^\tau C_{S_R}^{\tau\ast} \right) 
- 1.30\,|C_T^\tau|^2\,.
\end{align}
This is to be compared with 
\begin{align}
\frac{\mathcal{R}(\Lambda_c)}{\mathcal{R}_{\rm SM}(\Lambda_c)}\simeq
0.262\, \frac{\mathcal{R}(D)}{\mathcal{R}_{\rm SM}(D)} + 0.738\,\frac{\mathcal{R}(D^\ast)}{\mathcal{R}_{\rm SM}(D^\ast)}\,,
\end{align}
found in Ref.~\cite{Blanke:2019qrx}. As expected, the coefficients in the sum rule slightly change with this update due to the improvements in the form factors and the different choice employed relative to the minimization procedure for $\delta_{\Lambda_c}$. However, the change is only minimal, showing the robustness of the sum rule.

It is interesting to notice that a deviation in $\mathcal{R}(D^{\ast})$ from the SM has a stronger impact on   $\mathcal{R}(\Lambda_c)$ compared to one in $\mathcal{R}(D)$. Therefore, the latest measurement of LHCb~\cite{LHCbSem} with a value for $\mathcal{R}(D^{\ast})$ quite close to the SM value while that of $\mathcal{R}(D)$ being further away, decreases the expected deviation in $\mathcal{R}(\Lambda_c)$. 

Equation~\eqref{eq:Sumrule} holds in any tau-philic NP scenario described by the effective Hamiltonian in Eq.~\eqref{Heff}. Moreover, for $|C_T^\tau| \ll 1$, the correction factor $\delta_{\Lambda_c}$ is irrelevant. We therefore obtain the model-independent prediction 
\beq \label{eq:sumrule}
\mathcal{R}(\Lambda_c) &\simeq \mathcal{R}_{\rm SM}(\Lambda_c)
\left(
0.280\,
\frac{\mathcal{R}(D)}{\mathcal{R}_{\rm SM}(D)} + 
0.720\,
\frac{\mathcal{R}(D^\ast)}{\mathcal{R}_{\rm SM}(D^\ast)}
\right)\nonumber \\
 & = \mathcal{R}_{\rm SM}(\Lambda_c) \left( 1.172 \pm  0.038 \right) \nonumber  \\
&= 0.380 \pm 0.012 \pm 0.005\,,
\eeq
where the first error arises from the experimental uncertainty of $\mathcal{R}(D)$ and $\mathcal{R}(D^\ast)$, and the second one from $\mathcal{R}_{\rm SM}(\Lambda_c)$. Even with the updated coefficients, the central value of the predicted $\mathcal{R}(\Lambda_c)$ is unchanged with respect to Ref.~\cite{Blanke:2019qrx}, where $\mathcal{R}(\Lambda_c)=0.38\pm0.01\pm0.01$ was obtained; the only difference is in the increased precision of the prediction, more accurate due to the updates in the measurements of $\mathcal{R}(D^{(*)})/\mathcal{R}_{\rm SM}(D^{(*)})$ and once again proving the stability and robustness of the sum rule.

\section{NP Scenarios}\label{sec:NP}

In this Section we consider NP scenarios including WCs by the addition of at most two new heavy fields. In the cases of two fields, we allow one of them to couple to tau leptons, and the other to both light leptons with the same strength.\footnote{Strictly speaking, if the NP field coupling to light leptons is a leptoquark, the stringent constraints from lepton flavor violating decays require the introduction of \textit{two} such fields, one coupling to muons and the other to electrons~\cite{Crivellin:2022mff}.} We consider all NP WCs to be real, unless stated otherwise.

\subsection{Scalar Leptoquarks}

Out of the five scalar Leptoquarks (LQs)~\cite{Buchmuller:1986zs}, only three can generate $b\to cl\nu$ transitions:
\begin{itemize}
\item $S_1=(\mathbf{\bar{3}},\mathbf{1},1/3)$, an $SU(2)_L$-singlet scalar that, once integrated out, contributes to $C_{V_L}^l$ and/or the combination $C_{S_L}^l=-4C_T^l$ (at the matching scale), which becomes $C_{S_L}^l (\mu_b)\simeq - 8.9 C_T^l (\mu_b)$ at the decay scale. Solutions to the ${\cal R}(D)$ and ${\cal R}(D^*)$ anomalies by means of this LQ can be found in Refs.~\cite{
Sakaki:2013bfa,Freytsis:2015qca,Bauer:2015knc,Crivellin:2017zlb,Cai:2017wry,Altmannshofer:2017poe,Marzocca:2018wcf,Iguro:2018vqb,
Crivellin:2019qnh}. Note however that the $SU(2)_L$ symmetry implies an inevitable correlation between $C_{V_L}^l$ and a tree-level contribution to $b \to s \nu_l \overline{\nu}_l$. Hence, a constraint from $B\to K^*\nu\bar{\nu}$ measurement is unavoidable~\cite{Belle:2017oht,Carvunis:2021dss,Endo:2021lhi}. Moreover, additional severe bounds come from the $S_1$--$\nu_l$ box diagrams contributions to $\Delta M_s$~\cite{Crivellin:2019dwb,Cornella:2021sby}.
\item $R_2=(\mathbf{3},\mathbf{2},7/6)$, a weak doublet scalar whose footprints at the $B$-meson scale are described by a contribution satisfying $C_{S_L}^l=4C_T^l$. Once again, due to RGE this relation becomes $C_{S_L}^l (\mu_b) \simeq 8.4 C_T^l (\mu_b)$ at the low scale. In this specific scenario we allow the WCs to be complex, since this is a necessary requirement in order to address at the same time ${\cal R}(D)$ and ${\cal R}(D^*)$~\cite{Blanke:2018yud,Blanke:2019qrx,Becirevic:2016yqi,Becirevic:2018afm,Iguro:2018vqb,Iguro:2020keo,Crivellin:2022mff,Becirevic:2022tsj}.
\item $S_3=(\mathbf{\bar{3}},\mathbf{3},1/3)$, an $SU(2)_L$-triplet scalar that is parametrized at the low scale by the WC $C_{V_L}^l$. Models which contain such a solution to the LFU ratios have been studied in Refs.~\cite{Deshpande:2012rr,Tanaka:2012nw,Sakaki:2013bfa,Freytsis:2015qca,Crivellin:2017zlb,Marzocca:2018wcf,Becirevic:2022tsj}. Similarly to the $S_1$ case, also this scenario suffers from the constraint induced by $B\to K^*\nu\bar{\nu}$, severely limiting the allowed size for $C_{V_L}^l$~\cite{Sakaki:2013bfa}.
\end{itemize}
It is important to remember that, in order to avoid the undesirable effect of proton decays, one has to forbid di-quark couplings to the LQ for $S_1$ and $S_3$ (e.g. by a symmetry, see Ref.~\cite{Dorsner:2016wpm}).

\subsection{Vector Leptoquarks}

A second family of solutions for the LFU ratios involves vector LQs.\footnote{It is worth mentioning that these solutions usually require some sort of UV completion in order to explain the origin of a massive spin-1 particle.} Out of the five vector LQs~\cite{Buchmuller:1986zs} only two can produce effects of interest in our study, namely:
\begin{itemize}
\item $U_1=(\mathbf{3},\mathbf{1},2/3)$, an $SU(2)_L$-singlet vector that produces at the low scale the WCs $C_{V_L}^l$ and/or $C_{S_R}^l$. Models that include this LQ in a Pati-Salam extension of the SM can be found, e.g., in Refs.~\cite{Assad:2017iib,DiLuzio:2017vat,Calibbi:2017qbu,Bordone:2017bld,Barbieri:2017tuq,
Blanke:2018sro,Greljo:2018tuh,Bordone:2018nbg,
Crivellin:2018yvo,DiLuzio:2018zxy,
Cornella:2019hct,Iguro:2021kdw,
Cornella:2021sby,Iguro:2022ozl
}, which are based on the early studies performed in Refs.~\cite{Alonso:2015sja,Calibbi:2015kma,Fajfer:2015ycq,Barbieri:2015yvd,Hiller:2016kry,Bhattacharya:2016mcc,Barbieri:2016las,Buttazzo:2017ixm}.
\item $V_2=(\mathbf{\bar{3}},\mathbf{2},5/6)$, a weak doublet vector whose effects at the decay scale can be described by means of the WC $C_{S_R}^l$. An example for this kind of solution can be found, e.g., in Ref.~\cite{Cheung:2022zsb}. This scenario, previously disfavored due to its limited impact on $\mathcal{R}(D^*)$, is now viable again due to the recent LHCb result hinting at a smaller deviation in $\mathcal{R}(D^*)$ compared to the one in $\mathcal{R}(D)$~\cite{LHCbSem}. Note that, in order to avoid proton decay, also this scenario requires a symmetry that prevents di-quark coupling to $V_2$.
\end{itemize}

\subsection{Charged Higgses}

A charged scalar boson ($H^\pm$) generates the WCs $C_{S_R}^l$ and $C_{S_L}^l$. The 2HDM model of type II at large $\tan\beta$~\cite{Kalinowski:1990ba,Hou:1992sy,Nierste:2008qe} leads to the wrong sign to fit data, but the 2HDM with a generic flavor structure~\cite{Crivellin:2012ye,Celis:2012dk,Ko:2012sv,Crivellin:2013wna,Crivellin:2015hha,
Chen:2017eby,
Iguro:2017ysu,
Iguro:2022uzz,Blanke:2022pjy}, can lead to constructive effects. It is interesting to note that while a fit including only $C_{S_L}^l$ requires it to be complex in order to properly address the data, this is no longer necessary once both WCs are allowed at the same time, as in our fits.

\subsection{Singly charged vector boson} 

$W'$, being a charged vector boson, generates $C_{V_L}^l$~\cite{He:2012zp,Greljo:2015mma,Boucenna:2016wpr,He:2017bft,Abdullah:2018ets,Gomez:2019xfw}. However, such solutions are no more viable due to constraints from $\Delta M_s$, $b\to s\nu\nu$ and LHC direct searches like $pp (b\bar{b}) \to Z^\prime\to \tau^+\tau^-$, which arose due to $SU(2)_L$ invariance. Similarly, a $W_R^\prime$ scenario~\cite{Asadi:2018wea} is no longer compatible with collider bounds~\cite{IguroKEK}.

\section{Results and Discussion}\label{sec:res}

We are now ready to assess how 1D and 2D extension of the SM perform in explaining ${\cal R}(D)$, ${\cal R}(D^*)$ and ${\cal R}(\Lambda_c)$, where the dimensionality of the extension refers to the number of new fields, not to the number of WCs generated. For the form factors, we utilized results obtained by lattice QCD whenever available~\cite{MILC:2015uhg,FermilabLattice:2021cdg,Detmold:2015aaa}, therefore relying on the results of Heavy Quark Effective Theory~\cite{Bernlochner:2017jka} only for tensor form factors in $B \to D^*$ transitions. It is important to stress that our main conclusions are unchanged if one employs a different choice for the form factors. Our fits are performed by carrying out a Markov Chain Monte Carlo Bayesian analysis, implementing the full analytic expressions and including all theoretical correlations for all the analysed modes, using the \texttt{HEPfit} code~\cite{DeBlas:2019ehy}. We assumed for each WC a flat prior centered around zero, representing our initial ignorance on their allowed regions. To ascertain the prior-independence of our results and avoid misinterpretations, we ensured that the employed ranges for the priors were large enough, in such a way that the posterior distributions would not be cut (as an artifact that could be induced by being too constrictive without a motivated theoretical reason with the prior ranges), nor it would be altered by further enlarging such ranges. Given that our goal is to test whether the validity of the sum rule among the three LFU ratios at Eq.~\eqref{eq:sumrule} holds for any NP scenario, or if there are indeed scenarios in which either the correction factor $\delta_{\Lambda_c}$ defined in Eq.~\eqref{eq:deltaLambdac} is maximized or the sum rule is even violated by NP coupling to muons and electrons as well, a case not covered by our sum rule in Eqs.~\eqref{eq:Sumrule} and~\eqref{eq:deltaLambdac}, we proceed in the following way: as a first step and for the sake of simplicity, we perform a fit only to the ratios in order to assess how they comply with the sum rule; only in the case of a positive result, we therefore inspect and comment on how they fare once additional constraints are considered, like, e.g., the $B_c\to\tau\nu$ decay (which we allow to be as large as 60\%~\cite{Blanke:2018yud}), the $D^{*-}$ polarization~\cite{Belle:2019ewo} or constraints on $|V_{cb}|$ coming from fits to the Unitary Triangle~\cite{UTfit:2022hsi,CKMfitter}.

\begin{table}[!t!]
\centering
\newcommand{\bhline}[1]{\noalign{\hrule height #1}}
\renewcommand{\arraystretch}{1.4}
\begin{tabular}{c|cc|c}
\hline
\quad scenario~\, \quad & $\mathcal{R}(D)$ & $\mathcal{R}(D^*)$ & $\mathcal{R}(\Lambda_c)$\\
\rowcolor{gray!15}
exp. & 0.36(3) & 0.29(1) & 0.24(7) \\
\mbox{\boldmath{$S_1$}} &
$0.36 (3)$ & $0.29 (1)$ & $0.38 (3)$ \\ 
\mbox{\boldmath{$R_2$}} &
$0.36 (3)$ & $0.28 (1)$ & $0.40 (4)$ \\ 
\mbox{\boldmath{$S_3$}} &
$0.33 (2)$ & $0.29 (1)$ & $0.38 (2)$ \\ 
\mbox{\boldmath{$U_1$}}&
$0.36 (3)$ & $0.28 (1)$ & $0.37 (2)$ \\
\mbox{\boldmath{$V_2$}}&
$0.36 (3)$ & $0.28 (1)$ & $0.36 (1)$ \\
\mbox{\boldmath{$H^\pm$}} &
$0.36 (3)$ & $0.28 (1)$ & $0.36 (2)$ \\ 
\hline
\end{tabular}
\caption{\label{tab:RL}Predicted values for ${\cal R}(\Lambda_c)$ from a fit to ${\cal R}(D)$ and ${\cal R}(D^*)$ for several single particle extensions of the SM which couple to tau leptons.
}
\end{table}

\subsection{1D scenarios}

Assuming that NP couples to taus only, none of the extensions discussed in the previous Section is capable of describing in a satisfactory way the measurements of the three LFU ratios. This does not come as a surprise, since it was already known that the three ratios are connected by the sum rule derived in Refs.~\cite{Blanke:2018yud,Blanke:2019qrx} and Sec.\ \ref{sec:sumrule}: if two of them are measured above the SM prediction, like is the case for ${\cal R}(D)$ and ${\cal R}(D^*)$, a similar behavior is expected for the third one, contrary to what is the case for ${\cal R}(\Lambda_c)$. For a recent result of 1D global fits to all relevant data in this sector, we refer the reader to Ref.~\cite{Iguro:2022yzr}.

To better assess the (in)compatibility of data under the 1D hypotheses, we therefore performed the following test: first, for each scenario capable to fit ${\cal R}(D)$ and ${\cal R}(D^*)$, we predict the value for ${\cal R}(\Lambda_c)$; in a similar fashion, for each scenario capable to fit ${\cal R}(\Lambda_c)$, we predict the values for ${\cal R}(D)$ and ${\cal R}(D^*)$. It is worth mentioning that, as already observed in Refs.~\cite{Blanke:2018yud,Blanke:2019qrx}, the only prediction for ${\cal R}(\Lambda_c)$ affected (albeit marginally) by allowing the $B_c\to\tau\nu$ decay up to 60\% instead of a lower value, e.g.~30\%, is the one involving a charged Higgs. We report our findings in Tables~\ref{tab:RL} and~\ref{tab:RD}, respectively. Note that in Table~\ref{tab:RD} we report only the scenarios of a scalar LQ $S_1$ or $S_3$, or of a vector LQ $U_1$, since those are the only ones capable to reproduce the measured value of $\mathcal{R}(\Lambda_c)$. As expected, in the case where the mesons LFU ratios are considered in the fit, a large prediction for the baryon one is obtained, compatible with the prediction of the sum rule in Eq.~\eqref{eq:sumrule}, larger than the SM prediction and hence $\sim 2\sigma$ above its measured value. On the other hand, when predicting the values for ${\cal R}(D)$ and ${\cal R}(D^*)$ from a fit to ${\cal R}(\Lambda_c)$, the opposite pattern is observed: a value for the latter ratio complying with data would imply values for the former ones smaller than their SM predictions, and $\sim 2\sigma$ below their measured values. It is worth mentioning that, if one uses for ${\cal R}(\Lambda_c)$ the value suggested in Ref.~\cite{Bernlochner:2022hyz}, the discrepancy among predicted values and measured ones is reduced to $\sim 1.5\sigma$, as shown in Table~\ref{tab:RD_2}. 

Nevertheless, the current uncertainty on ${\cal R}(\Lambda_c)$ is still large enough that those scenarios cannot be ruled out at present, and a potential decrease in the discrepancy among ${\cal R}(D)$, ${\cal R}(D^*)$ and their SM prediction could reduce the induced tension in ${\cal R}(\Lambda_c)$, or vice versa.

\begin{table}[!t!]
\centering
\newcommand{\bhline}[1]{\noalign{\hrule height #1}}
\renewcommand{\arraystretch}{1.4}
 \begin{tabular}{c|c|cc}
\hline
\quad scenario~\, \quad & $\mathcal{R}(\Lambda_c)$  & $\mathcal{R}(D)$ & $\mathcal{R}(D^*)$\\
\rowcolor{gray!15}
exp. & 0.24(7) & 0.36(3) & 0.29(1) \\
 \mbox{\boldmath{$S_1$}} &
$0.23 (7)$ & $0.21 (8)$ & $0.16 (8)$ \\ 
\mbox{\boldmath{$S_3$}} &
$0.21 (8)$ & $0.18 (7)$ & $0.17 (6)$ \\ 
\mbox{\boldmath{$U_1$}}&
$0.22 (8)$ & $0.15 (8)$ & $0.17 (8)$ \\
\hline
\end{tabular}
\caption{\label{tab:RD}Predicted values for ${\cal R}(D)$ and ${\cal R}(D^*)$ from a fit to the experimental value of ${\cal R}(\Lambda_c)$~\cite{LHCb:2022piu}.}
\end{table}
\begin{table}[!t!]
\centering
\newcommand{\bhline}[1]{\noalign{\hrule height #1}}
\renewcommand{\arraystretch}{1.4}
\begin{tabular}{c|c|cc}
\hline
\quad scenario~\, \quad & $\mathcal{R}(\Lambda_c)$  & $\mathcal{R}(D)$ & $\mathcal{R}(D^*)$\\
\rowcolor{gray!15}
Ref.~\cite{Bernlochner:2022hyz} & 0.29(7) & 0.36(3) & 0.29(1) \\
\mbox{\boldmath{$S_1$}} &
$0.28 (7)$ & $0.25 (8)$ & $0.19 (8)$ \\ 
\mbox{\boldmath{$S_3$}} &
$0.27 (7)$ & $0.23 (6)$ & $0.21 (6)$ \\ 
\mbox{\boldmath{$U_1$}}&
$0.28 (7)$ & $0.17 (9)$ & $0.22 (8)$ \\
\hline
\end{tabular}
\caption{\label{tab:RD_2}Predicted values for ${\cal R}(D)$ and ${\cal R}(D^*)$ using the value of ${\cal R}(\Lambda_c)$ from Ref.~\cite{Bernlochner:2022hyz}, assuming $|V_{cb}|=0.04$.}
\end{table}

\subsection{2D scenarios}

Here we allow a first new field to couple to taus only, and a second one to couple to muons and electrons equally. For this reason, we identify fields belonging to the first class with the label $\tau$, e.g., $R_2^\tau$, while fields related to the second one are labelled with $\ell$, e.g.,  $S_1^\ell$. Having at hand six  possible kinds of fields parametrized by a different low-energy EFT description, and  each being allowed to couple either to the heavy charged lepton or to the light ones, we ultimately inspected a total of 36 potential 2D scenarios. Out of all these possibilities, we only found two scenarios capable to reproduce in a satisfactory way all three LFU ratios, with all the other scenarios still implying an over-production of taus in ${\cal R}(\Lambda_c)$ at the $2\,\sigma$ level. The first viable model is composed by an $S_1$ LQ coupling to light fermions, together with an $R_2$ coupled to taus, namely the pair formed by $S_1^\ell$ and $R_2^\tau$. The second possibility shares the same NP extension coupled to muons and electrons, but requires furthermore a charged Higgs coupled to taus, i.e., the pair formed by $S_1^\ell$ and $H^{\pm\tau}$. The fact that both scenarios rely on the presence of an $SU(2)_L$-singlet scalar LQ coupled to light fermions is the reason why these scenarios  apparently comply with data, but is also the origin why they  ultimately fail once faced with additional constraints.

Indeed, once NP is allowed to couple to both heavy and light charged leptons, the numerical formulae for the LFU ratios and for the sum rule connecting them have to be modified accordingly. Observing now that $S_1^\ell$ implies the presence of a tensor WC, a strong violation of the sum rule (and hence a potential opposite behavior of ${\cal R}(\Lambda_c)$ w.r.t.\,${\cal R}(D)$ and ${\cal R}(D^*)$) could be induced in the case of a non-negligible size for this coefficient. This is indeed what we find in our fits, where in both viable scenarios we need a strong contribution to the scalar and tensor currents, equal to $C_{S_L}^\ell=-4C_T^\ell \simeq \pm 1$, in order to obtain a value ${\cal R}(\Lambda_c) \simeq 0.24$.

Moreover, this is not the only requirement for the WCs coupling to light leptons: $S_1^\ell$ also generates a vector current mediated by $C_{V_L}^\ell$, whose value is constrained by the fit in strong correlation with that of the former pair of WCs, and determined to be in both scenarios equal to $C_{V_L}^\ell\simeq -1$. This corresponds to a $-100\%$ correction in the light leptons vector current w.r.t.\,the SM contribution, inducing a complete cancellation of this term.

However, these solutions are actually not viable once further constraints are taken into account. NP contributions to vector currents involving light leptons are constrained by high-$p_{\rm T}$ lepton tail searches at the LHC $|C_{V_L}^{e}|<0.25$ \cite{Iguro:2020keo,Allwicher:2022mcg} and, even more, by the aforementioned bound imposed on $S_1$ LQs by $B\to K^*\nu\bar{\nu}$ measurement~\cite{Belle:2017oht}, which implies $-0.011\le C_{V_L}^\ell\le 0.027$~\cite{Endo:2021lhi}; moreover, a strong NP tensor component for light leptons is also heavily constrained by the high-$p_{\rm T}$ search \cite{ATLAS:2019lsy}, which implies $|C_T^{e}|< 0.32$, or by an analysis of angular distribution~\cite{Belle:2017rcc,Belle:2018ezy} and $D^{*-}$ polarization data~\cite{Belle:2019ewo}, which requires it to be even smaller, namely  $|C_T^\ell|\le 0.05$~\cite{Jung:2018lfu,Iguro:2020cpg}. NP coupling to light leptons is also strongly constrained by global CKM fits~\cite{UTfit:2022hsi,CKMfitter}, with the value extracted for $|V_{cb}|$ from $B \to D^{(*)} \ell \nu$ being consistent with the constraint coming from a fit to all the other channels entering in the Unitarity Triangle Analysis (UTA). Since a modification to such effective couplings would unavoidably alter this extraction, stringent bounds can therefore be inferred on the NP WCs, see Ref.~\cite{Jung:2018lfu} for an early non-universal study. A systematic analysis of these bounds goes beyond the scope of this paper, given that the aforementioned constraints coming from angular distributions data, $D^{*-}$ polarization and collider bounds are already strong enough to invalidate the selected 2D scenarios apparently capable to describe the LFU ratios. However, a crude estimate of the bound for a WC can be obtained by requiring that the extracted value for $|V_{cb}|$ in the presence of NP would stay compatible with its predicted value by a UTA where such channels are not included in the fit~\cite{UTfit:2022hsi,CKMfitter}. Such estimates lead to $|C_{V_L}^{\ell}| \lesssim 0.025$ and $|C_T^{\ell}| \lesssim 0.25$.

\subsection{General Model-Independent fit}

For completeness, we conclude our analysis of viable NP scenarios by performing a fully model-independent fit for eight generic WCs, i.e., $C_{V_L,S_{L,R},T}^{\tau,\ell} \neq 0$, which we take to be real.\footnote{We do not consider the introduction of 
imaginary parts for these coefficients 
to be able to alter our conclusions, mainly because the interference terms among different WCs in the sum rule of Eq.~\eqref{eq:sumrule} are proportional to lepton masses, hence negligible for $\ell=e,\,\mu$. Similarly, since for right-handed neutrino interactions interference terms do not exist for any of the three modes, their inclusion would not have an impact on our results
.} The results turn out to be similar to the ones observed for the 2D scenarios: while it is indeed possible to find regions of the eight-dimensional WC parameter space where the values for all three LFU ratios are found to be compatible with observed measurements, when additional constraints like the $|V_{cb}|$ determination within CKM fits, angular distributions data, $D^{*-}$ polarization and collider bounds are considered, these solutions are no longer acceptable.

\section{Conclusions}\label{sec:concl}

In this paper, we have critically analysed the latest results concerning LFU ratios in $B$-meson charged-current decays, aiming to assess the compatibility of data which is challenged by the measurement of ${\cal R}(\Lambda_c)$: the latter result, being smaller than the SM prediction, is in contrast with ${\cal R}(D)$ and ${\cal R}(D^*)$, where an enhancement w.r.t.~the SM of $3.2\,\sigma$ is observed. Since all these ratios are mediated by the same $b\to c l \bar\nu$ transition, their NP predictions are connected in a model-independent way by a sum rule, which we updated here while investigating whether the data at hand complies with it. For this we have relaxed one assumption of the sum rule, namely that NP affects $b\to c \tau\nu$ but not $b\to c \ell\nu$ with $\ell=e,\mu$. Due to the sum rule, no single particle can explain all three LFU ratios at the same time, and even when considering the different normalization suggested in Ref.~\cite{Bernlochner:2022hyz} for ${\cal R}(\Lambda_c)$ the discrepancy is still at the $\sim 1.5\,\sigma$ level.

We therefore investigated whether the addition of a second NP field, this time coupling equally to light charged leptons $\ell=e,\mu$, could induce a modification in the sum rule such that it is possible to address the opposite behavior of ${\cal R}(\Lambda_c)$ compared to ${\cal R}(D^{(*)})$. While we found two possible scenarios capable to address the three LFU ratios at the same time, namely one formed by the pair $S_1^\ell$ and $R_2^\tau$, and the second formed by the couple $S_1^\ell$ and $H^{\pm\tau}$, we ultimately ruled out these possibilities as well using CKM fits, $B\to K^*\nu\bar{\nu}$, angular distributions and high-$p_T$ collider bounds.

We further performed a fit to eight WCs, half of them related to taus with the remaining associated with light charged leptons. Even in this general case we found that while a fit to the three LFU ratios might find viable solutions in the eight-dimensional WC parameter space, once additional constraints are taken into account such a solution is no longer acceptable.  We therefore concluded that present data cannot be addressed, neither in the SM nor beyond, in a satisfactory way, as current experimental results for $\mathcal{R}(D^{(*)})$ and $\mathcal{R}(\Lambda_c)$ show an inconsistent pattern. It is therefore mandatory to obtain further experimental results in this sector in order to eventually converge to a coherent data pattern, differently from what we currently observe. Whether this pattern will lead us to the SM or to NP, only time will tell.

With our current input and assumptions, we predict that at least one of the central values of $\mathcal{R}(D^{(*)})$ or $\mathcal{R}(\Lambda_c)$ will move from its present value once more statistics is accumulated, independently of the presence or nature of NP. Moreover, it might also be possible that NP is present in the $q^2$ distributions of light lepton modes, while still resulting in consistent values for $|V_{cb}|$, if a different theoretical approach for the form factors is used~\cite{Martinelli:2022xir}. Interestingly, this might provide a connection to the anomaly in $\Delta A_{FB}$~\cite{Bobeth:2021lya}, which requires different NP related to muons and electrons~\cite{Carvunis:2021dss,Bhattacharya:2022bdk}. Furthermore, in a UV complete (or simplified) model, NP effects in $\Delta F=2$ processes occur in general, such that the global CKM fit could allow for larger NP effects in the determination of $|V_{cb}|$.

\vspace{2mm} {\it Acknowledgments.}--- {\small The work of M.B., M.F., S.I. and U.N. is supported by the Deutsche Forschungsgemeinschaft (DFG, German Research Foundation) under grant 396021762-TRR 257. A.C. is supported by a Professorship Grant (PP00P2\_176884) of the Swiss National Science Foundation. The work of T.K. is supported by the Japan Society for the Promotion of Science (JSPS) Grant-in-Aid for Early-Career Scientists (Grant No. 19K14706) and the JSPS Core-to-Core Program (Grant No. JPJSCCA20200002). R.W. is partially supported by the INFN grant ‘FLAVOR’ and the PRIN 2017L5W2PT.}

\bibliography{BIB}

\end{document}